\documentclass[letterpaper,prb,twocolumn,showpacs]{revtex4}
\usepackage{graphicx,bm}
\usepackage{color}
\begin{document}
\newcommand{\mybm}[1]{\mbox{\boldmath$#1$}}
\newcommand{\mysw}[1]{\scriptscriptstyle #1}

\title{Ferromagnetism in Armchair Graphene Nanoribbon}

\author{Hsiu-Hau Lin}
\email{hsiuhau@phys.nthu.edu.tw}
\affiliation{Department of Physics, National Tsing-Hua University, Hsinchu 300, Taiwan}
\affiliation{Physics Division, National Center for Theoretical Sciences, Hsinchu 300, Taiwan}

\author{Toshiya Hikihara}
\email{hikihara@phys.sci.hokudai.ac.jp}
\affiliation{Department of Physics, Hokkaido University, Sapporo 060-0810, Japan}

\author{Horng-Tay Jeng}
\email{jeng@phys.sinica.edu.tw}
\affiliation{Institute of Physics, Academia Sinica, Nankang, Taipei 11529, Taiwan}
\affiliation{Department of Physics, National Tsing-Hua University, Hsinchu 300, Taiwan}

\author{Bor-Luen Huang}
\affiliation{Department of Physics, National Tsing-Hua University, Hsinchu 300, Taiwan}

\author{Chung-Yu Mou}
\affiliation{Department of Physics, National Tsing-Hua University, Hsinchu 300, Taiwan}

\author{Xiao Hu}
\affiliation{WPI Center for Materials Nanoarchitectonics, National Institute for Materials Science, Tsukuba 305-0047, Japan}
\date{\today}
\begin{abstract}
Due to the weak spin-orbit interaction and the peculiar relativistic dispersion in graphene, there are exciting proposals to build spin qubits in graphene nanoribbons with armchair boundaries. However, the mutual interactions between electrons are neglected in most studies so far and thus motivate us to investigate the role of electronic correlations in armchair graphene nanoribbon by both analytical and numerical methods. Here we show that the inclusion of mutual repulsions leads to drastic changes and the ground state turns ferromagnetic in a range of carrier concentrations. Our findings highlight the crucial importance of the electron-electron interaction and its subtle interplay with boundary topology in graphene nanoribbons. Furthermore, since the ferromagnetic properties sensitively depends on the carrier concentration, it can be manipulated at ease by electric gates. The resultant ferromagnetic state with metallic conductivity is not only surprising from an academic viewpoint, but also has potential applications in spintronics at nanoscale.
\end{abstract}
\maketitle

\section{Introduction}

Graphene,\cite{Geim07} a single hexagonal layer of carbon atoms in two dimensions (2D), is the building block for graphitic materials ranging from 0D fullerenes to 1D nanotubes, and also the commonly seen 3D graphite. Since it was generally believed that the two-dimensional lattice should not exist at finite temperature, graphene had often been used as a toy model and viewed as an academic material until its recent discovery in laboratory.\cite{Novoselov04} The honeycomb structure gives rise to linear dispersion, making electrons and holes in graphene massless as in relativistic theories.\cite{Semenoff84,Haldane88} Therefore, most studies focus on the electronic and transport properties arisen from this peculiar band structure,\cite{Novoselov05,Zhang05,Berger06,Ohta06,Bostwick07} such as the half-integer quantum Hall effect\cite{Novoselov05,Zhang05} due to the $\pi$ Berry phase, the quantization of minimal conductivity where carriers are almost absent\cite{Novoselov05} and so on.

One of the potential applications of graphene is to realize fast electronics at nanoscale, making graphene nanoribbons (GNRs) a natural building block for these devices. Since the electronic structure sensitively depends on the transverse width and also the edge topology, there are intensive investigations\cite{Son06,Barone06,Areshkin07,Liang07,Nakada96} on narrow GNRs with width less than 10 nm. Although GNRs have been successfully fabricated by lithography\cite{Han07} down to the widths of 20 nm, the roughness of the edges remains large (about 5 nm or larger). As a result, theoretical predictions may not be applicable and limit the fundamental and practical applications. A recent breakthrough of fabricating GNRs came from chemical methods.\cite{Li08} It is rather remarkable that the width of the GNRs can be fabricated in a controlled way down to 10 nm. In addition, the edges of these GNRs are ultra smooth with possibly well-defined zigzag or armchair shapes, suitable for building electric junctions at molecular scale.

As the transverse width shrinks, the quantum fluctuations become important and results/predictions from mean-field theories shall be checked carefully. Meanwhile, since the open boundaries of GNR play a crucial role at nanoscale, the interplay between the Coulomb interaction and the edge morphology will lead to rich physics. For instance, it has been revealed that the Coulomb interaction gives rise to edge moments in zigzag GNR.\cite{Fujita96,Okada01,Hikihara03} Furthermore, Son, Cohen, and Louie\cite{Son06b} have shown that, in the presence of external electric field in the transverse direction, the system turns half metallic with magnetic properties controlled by the external electric field. Their results not only show that the electronic spin degrees of freedom can be manipulated by the electric fields, but also bring up the possibility to explore spintronics\cite{Wolf01,DasSarma04,MacDonald05} at the nanometer scale based on graphene.

\begin{figure}
\centering
\includegraphics[width=\columnwidth]{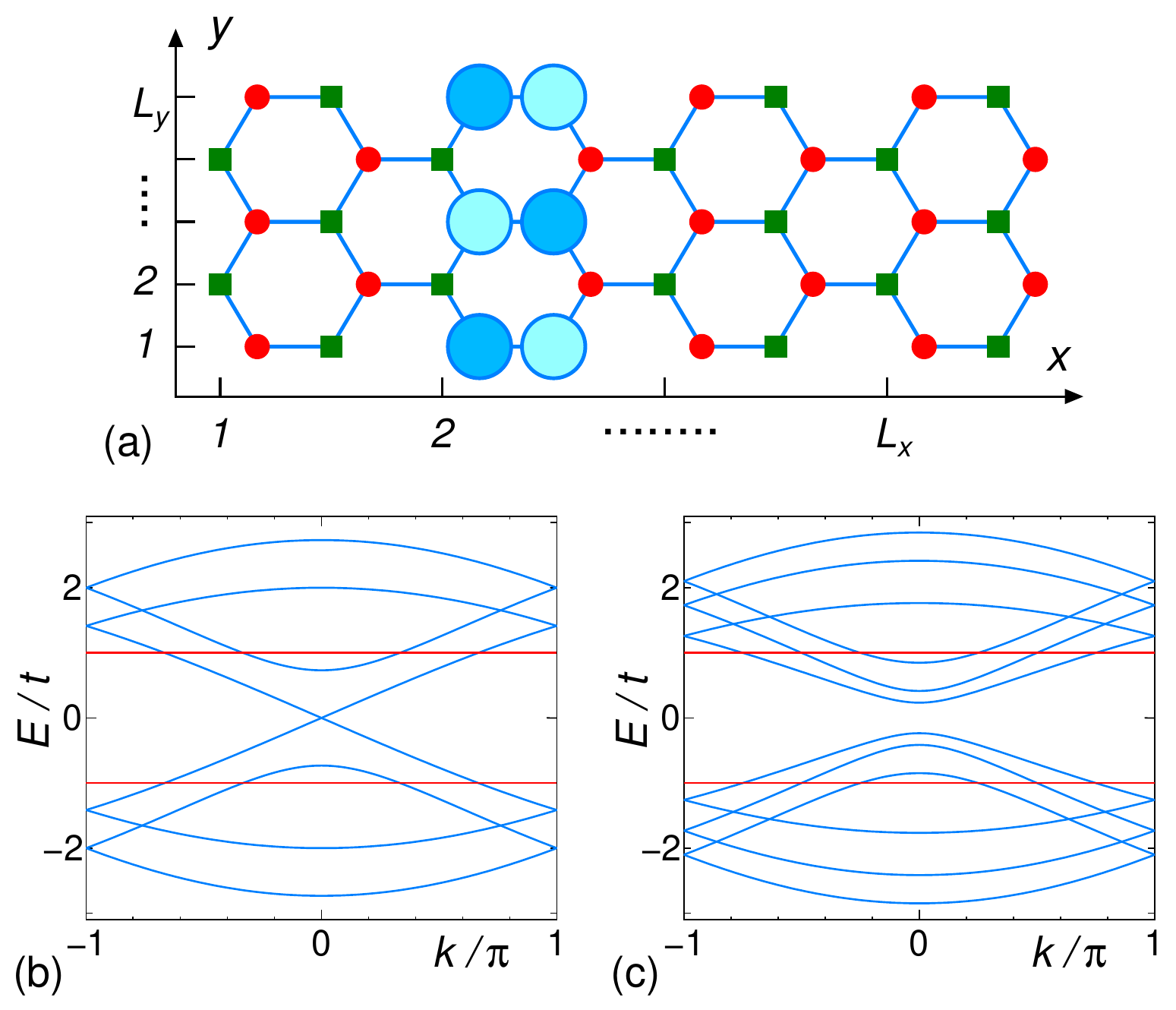}
\caption{\label{fig:Wannier} (Color online) (a) Armchair GNR of $L_{x}=4$ and $L_{y}=5$. Open edges are present at $y=1$ and $L_{y}$. The solid circles and squares represent sublattice A and B respectively. The shaded circles show the amplitudes of a local Wannier orbital of $E=t$ at $x=2$, with opposite signs indicated by light/dark colors. Band structures for the infinitely long GNR with (b) $L_y=5$ and (c) $L_y=7$ clearly show the pair of flat bands at $E = \pm t$.}
\end{figure}

Inspired by these discoveries, we investigate the effect of Coulomb interaction in armchair GNR as schematically shown in Fig.~\ref{fig:Wannier}. Note that the edges are hydrogenated so that the dangling $\sigma$ bonds are saturated and only the $\pi$ bands remain active in low energy.\cite{Son06b} By combining analytical weak-coupling analysis, numerical density matrix renormalization-group (DMRG) method, and the first-principles calculations, we show that armchair GNR exhibits an interesting carrier-mediated ferromagnetism upon appropriate doping. Even though only $\pi$ bands are active in low-energy, in appropriate doping regimes, the armchair edges give rise to both itinerant Bloch and localized Wannier orbitals. As will be explained later, these localized orbitals are direct consequences of quantum interferences in armchair GNR and form flat bands with zero velocity. The carrier-mediated ferromagnetism can thus be understood in two steps: Electronic correlations in the flat band generate intrinsic magnetic moments first, then the itinerant Bloch electrons mediate ferromagnetic exchange coupling among them. As a result, the magnetic properties of armchair GNR sensitively depend on the doping and thus can be manipulated easily by the external electric fields.

Though the ferromagnetic state in armchair GNR stems from the flat-band states which are partially filled, the mechanism is different from the ^^ ^^ flat-band ferromagnetism" proposed by Mielke and Tasaki.\cite{Mielke93,Tasaki98,Tanaka07} The key to Mielke-Tasaki ferromagnetism is the finite overlap of adjacent Wannier orbitals in the flat band: the finite overlaps generate exchange coupling among these orbitals and lead to the flat-band ferromagnetism. However, the Wannier orbital in armchair GNR (shown in Fig.~\ref{fig:Wannier}) has {\em zero} overlap with its adjacent neighbors. The flat band alone only accounts for the existence of the magnetic moments and the ferromagnetic order sets in only when itinerant carriers are present. A similar mechanism of ferromagnetism has been argued for the Hubbard model in a kind of frustrated lattice.\cite{TanakaI98,TanakaFM}

In the following, we will elaborate in details how the carrier-mediated ferromagnetism emerges (upon appropriate doping) in the armchair GNR. In Sec.\ \ref{sec:model}, we start with the Hubbard model and solve the band structure by the method of generalized Bloch theorem. In Sec.\ \ref{sec:weakcoupling}, we integrate out the gapped modes and explore the ground state properties in weak coupling. We first show how the local moments in the flat band form from the Coulomb interaction. We also demonstrate the crucial role of itinerant carriers in dispersive bands, which mediate the indirect exchange coupling among the local moments and give rise to the ferromagnetic ground state. Indeed, without those carriers, the ferromagnetism disappears and only Curie-like susceptibility remains in armchair GNR. In Sec.\ \ref{sec:DMRG}, we employ the technique of non-Abelian DMRG to investigate the ferromagnetic ground state in intermediate coupling. It is remarkable that the numerical results agree with the weak-coupling analysis pretty well. In Sec.\ \ref{sec:LSDA}, the realistic band structure and the long-range Coulomb interactions are included via the first-principles calculations. It is rather unexpected that the flat band is robust even when the realistic band structure is taken into account. Ferromagnetism appears around the same doping regime as predicted by either weak-coupling or DMRG approaches. Finally, in the last section, we discuss the robustness of our predictions and also their connections to practical experiments. 

\section{Hubbard Model for Armchair Nanoribbon}\label{sec:model}

To understand the carrier-mediated flat-band ferromagnetism in the armchair GNR, we start with the Hubbard Hamiltonian,
\begin{eqnarray}
H = -t \sum_{\langle{\bf r,r'}\rangle,\alpha} [c^{\dag}_{\alpha}({\bf r}) c^{}_{\alpha}({\bf r'}) + \mbox{H.c.}]
+ U \sum_{{\bf r}} n_{\uparrow}({\bf r}) n_{\downarrow}({\bf r}),
\end{eqnarray}
where $t$ is the hopping amplitude on the honeycomb network, $U>0$ is the on-site repulsion, $\alpha = \uparrow, \downarrow$ is the spin index.
The lattice points ${\bf r} = (x,y,\Lambda)$ are labeled by integer indices $(x,y)$ in longitudinal and transverse directions and the additional sublattice index $\Lambda = A, B$. The transverse integer index $y = 1,2,...,L_{y}$ defines the width of the ribbon while $x=1,2,...,L_x$ defines the length. The sum $\sum_{\langle{\bf r,r'}\rangle}$ is taken only for nearest-neighbor (NN) bonds. 

The values of $t$ reported in the literature\cite{Mintmire92,Wildoer98,Odom98} range from 2.4-2.7 eV for nanotubes, while $t \simeq 3$ eV is typical in graphites. Although an accurate value of $U$ is not yet known in GNRs, the estimate from polyacetylene, $U \simeq$ 6-10 eV,\cite{Baeriswyl85,Jeckelmann94} might serve as a reasonable guess. Thus, we expect the ratio $U/t$ to be of order one. 

Let us try to obtain the band structure of the armchair GNR within the tight-binding model first. For convenience, we consider the infinite length $L_x \to \infty$ in the following. Since the system is translational invariant along the $x$-direction, the hopping Hamiltonian simplifies after the partial Fourier transformation,
\begin{eqnarray}
c({\bf r}) = \frac{1}{\sqrt{L_{x}}} \sum_{k} e^{ik[x+\delta({\bf y})]} c(k,{\bf y}).
\label{Fourier}
\end{eqnarray}
We omit the spin index $\alpha$ for a while.
The shorthand notation ${\bf y} = (y,\Lambda)$ is defined to label the sites within a unit cell. The geometric phases arisen from the underlying honeycomb structure are
\begin{eqnarray}
\delta({\bf y}) = 2/3, &\qquad& \mbox{for $y$=even and $\Lambda=A$},
\nonumber\\
\delta({\bf y}) = 0, &\qquad& \mbox{for $y$=even and $\Lambda=B$},
\nonumber\\
\delta({\bf y}) = 1/6, &\qquad& \mbox{for $y$=odd and $\Lambda=A$},
\nonumber\\
\delta({\bf y}) = 1/2, &\qquad& \mbox{for $y$=odd and $\Lambda=B$.}
\end{eqnarray}
The hopping Hamiltonian simplifies to decoupled two-leg ladders with finite length $L_{y}$ labeled by momentum $k$ in the $x$-direction,
\begin{eqnarray}
H_{t} = \sum_{k} \Psi^{\dag}(k) H(k) \Psi(k),
\end{eqnarray}
where $\Psi(k)=[c(k,y;A), c(k,y;B)]$ is the fermion operator for sublattices $A$ and $B$ and thus has $2 L_{y}$ components. The reduced hopping Hamiltonian $H(k)$ is a $2 L_{y} \times 2 L_{y}$ matrix. It is rather interesting that $H(k)$ can be casted into supersymmetric (SUSY) form\cite{Mou04,Lin05,Zheng07}
\begin{eqnarray}
H(k) = \left( \begin{array}{cc}
0&Q^{\dag}\\
Q&0
\end{array}\right).
\label{SUSY}
\end{eqnarray}
The submatrix that connects opposite sublattices is the supercharge operator,
\begin{eqnarray}
Q = \left( \begin{array}{cccccc}
t_2^*&t_1&0&0&...&0\\
t_1&t_2^*&t_1&0&...&0\\
0&t_1&t_2^*&t_1&...&0\\
.&.&.&.&...&.\\
0&0&0&0&...&t_2^*
\end{array}\right),
\end{eqnarray}
where the complex hopping amplitudes are $t_1=-te^{ik/6}$ and $t_2=-te^{ik/3}$. One should not be surprised by the complex hopping amplitudes that come from the corresponding geometric phases $k \delta({\bf y})$.

To solve the wave function, let us introduce a $2L_y$-component spinor,
\begin{eqnarray}
\Phi({\bf y}) = \left[\begin{array}{c}
\varphi_{\mysw{A}}(y)\\
\varphi_{\mysw{B}}(y)
\end{array}\right],
\end{eqnarray}
where $\varphi_{\mysw{A/B}}(y)$ are the wave functions on sublattices A/B. Making use of the SUSY form in Eq.~(\ref{SUSY}), the coupled Harper equations are
\begin{eqnarray}
Q\: \varphi_{\mysw{A}} = E \varphi_{\mysw{B}},
\qquad
Q^{\dag} \varphi_{\mysw{B}} = E \varphi_{\mysw{A}}.
\end{eqnarray}
It is known that the solution for $E \neq 0$ is supersymmetric and can be solved by combining the two Harper equations together
$
Q^{\dag} Q\: \varphi_{\mysw{A}} = E^2 \varphi_{\mysw{A}}.
$ Once the eigenstate $\varphi_{\mysw{A}}$ is obtained, the wave function on the other sublattice is
$\varphi_{\mysw{B}} = \frac{1}{E} Q\: \varphi_{\mysw{A}}.$
Thus, the remaining task is to diagonalize the matrix $Q^\dag Q$. Note that the above trick only works for $E \neq 0$ eigenstates while the $E=0$ states can be obtained by finding the null space of $Q$ and $Q^\dag$ alone.

Before digging into details, we would like to make some general remarks. It is clear that, for each solution $\varphi_{\mysw{A}}$, we can construct two wave functions on the other sublattice $\varphi_{\mysw{B}}$ by two choices of eigenenergies $E = \pm |E|$. As a result, the energy spectrum is symmetric about $E=0$ and total wave functions of opposite energies $E = \pm |E|$ only differ by an overall minus sign on one of the sublattices (but not on the other). 

With this general picture in mind, we now turn to the explicit form of the matrix $Q^\dag Q$, that can be worked out by straightforward algebra
\begin{equation}
Q^\dag Q= \left( \begin{array}{cccccccc}
V_0-T_2&T_1&T_2&0&0&0&...&0\\
T_1&V_0&T_1&T_2&0&0&...&0\\
T_2&T_1&V_0&T_1&T_2&0&...&0\\
0&T_2&T_1&V_0&T_1&T_2&...&0\\
.&.&.&.&.&.&...&.\\
0&0&0&0&...&T_2&T_1&V_0-T_2\\
\end{array}\right),
\end{equation}
where $V_0=3t^2$, $T_1=2t^2 \cos(k/2)$ and $T_2 = t^2$. This matrix resembles the hopping Hamiltonian of a finite chain with the site potential $V_0$, nearest-neighbor hopping $T_1$ and next-nearest-neighbor one $T_2$. The eigenfunction satisfies
\begin{eqnarray}
&& \hspace{-5mm} T_{2}[\varphi_{\mysw{A}}(y+2)+\varphi_{\mysw{A}}(y-2)] + T_{1}[\varphi_{\mysw{A}}(y+1)+\varphi_{\mysw{A}}(y-1)]
\nonumber\\
&& \hspace{20mm} + V_0 \varphi_{\mysw{A}}(y) = E^2 \varphi_{\mysw{A}}(y),
\label{T1T2Hopping}
\end{eqnarray}
where $y=1,2,...,L_y$ with the appropriate boundary conditions
\begin{eqnarray}
\varphi_{\mysw{A}}(0) = 0, &&
\varphi_{\mysw{A}}(L_y+1) = 0,
\\
\varphi_{\mysw{A}}(-1) = - \varphi_{\mysw{A}}(1), &&
\varphi_{\mysw{A}}(L_y+2) = -\varphi_{\mysw{A}}(L_y).
\end{eqnarray}
The first two boundary conditions arise from the open ends of the effective two-leg ladder and the last two constraints comes from the change of the potential at the end sites. Note that the usual plane-wave solutions satisfy the bulk Harper equation in Eq.~(\ref{T1T2Hopping}). Thus, we only need to form appropriate linear combination of these plane-wave solutions to match the boundary conditions. In the case we study here, the eigenstate is the simple combination of opposite momentum states with a relative minus sign, i.e. the wave function takes the usual sine form,
\begin{eqnarray}
\varphi_{\mysw{A}}(y) = \sin(q_{m} y),
\end{eqnarray}
where the magnitude of transverse momentum is quantized, $q_m = m\pi/(L_{y}+1)$ with $m=1,2,...,L_y$. From the eigenstates, it is straightforward to compute the corresponding dispersions for each band,
\begin{eqnarray}
E &=& \pm [V_0+2T_1 \cos q_m + 2T_2 \cos 2q_m]^{1/2}
\nonumber\\
&=& \pm t [1+ 4 \cos(k/2) \cos q_m + 4 \cos^2 q_m]^{1/2}.
\label{Band}
\end{eqnarray}
With the wave function on sublattice $A$ and the energy dispersion, we can obtain the wave function on sublattice $B$ by the supercharge operator as described before. However, a closer look would ensure us that the full wave function on the armchair GNR is far simpler than we expected.

The simplification arises from the fact that the supercharges $Q$ and $Q^{\dag}$ commute. As a result, the matrix $Q^{\dag}Q = Q Q^{\dag}$ share the same eigenstates as the matrices $Q$ and $Q^\dag$. It is straightforward to show that the wave function $\varphi_{\mysw{A}}(y)$ is also an eigenstate of $Q$ with complex eigenvalue,
\begin{eqnarray}
Q \varphi_{\mysw{A}} = (t_2^*+ 2 t_1 \cos q_m) \varphi_{\mysw{A}} = |E| e^{i\theta({\bf k})} \varphi_{\mysw{A}},
\end{eqnarray}
where the phase $\theta({\bf k}) = \theta(k,q_m)$ of the complex eigenvalue is
\begin{eqnarray}
e^{i\theta({\bf k})} = \frac{t_2^*+2t_1 \cos q_m}{|E|}.
\end{eqnarray}
Therefore, the wave function on the sublattice $B$ is a duplicate of $\varphi_{\mysw{A}}$ with a momentum-dependent phase shift,
\begin{eqnarray}
\varphi_{\mysw{B}}(y) = \pm e^{i\theta({\bf k})} \sin(q_{m} y).
\end{eqnarray}
The $\pm$ signs arise from the signs of the energy, corresponding to antibonding and bonding bands respectively. Finally, after attaching appropriate renormalization factor, the full eigenstate wave function is labeled by the quantum number ${\bf k}=(k, q_m, s)$, including the longitudinal momentum $k$, the transverse momentum $q_m$ and antibonding/bonding index $s=\pm 1$. The explicit form of the wave function is
\begin{eqnarray}
\Phi_{\bf k}({\bf y}) = \frac{1}{\sqrt{L_y+1}} \left[
\begin{array}{c}
\sin (q_m y)\\
s e^{i\theta({\bf k})} \sin (q_m y),
\end{array}\right].
\end{eqnarray}
Note that the dependence of the longitudinal momentum $k$ only appears through the relative phase $\theta({\bf k})$ between wave functions on two sublattices. This simple analytical form of the eigenstates allows us to map the armchair GNR to effective theory in the low-energy limit and study the correlation effects analytically.

For the width of odd $L_y$, the transverse momentum goes through the particular value $q_m= \pi/2$, rendering the energy dispersion completely flat at energy $E=\pm t$ in the whole Brillouin zone. To obtain the wave function, we only need to evaluate the phase shift, $e^{i\theta_{\mysw{F}\pm}} = [t_2^*+ 2 t_1\cos (\pi/2)]/t = - e^{-ik/3}.$
Therefore, the wave functions for the flat bands $E= \pm t$ are 
\begin{eqnarray}
\Phi_{\mysw{F}\pm}({\bf y}) = \frac{1}{\sqrt{L_y+1}} \left[
\begin{array}{c}
\sin (\pi y/2)\\
\mp e^{-ik/3} \sin (\pi y/2)
\end{array}\right].
\end{eqnarray}
Since all states with different momentum $k$ are exactly degenerate, it is possible to construct the local Wannier orbital with the same energy,
\begin{eqnarray}
\Psi_{\mysw{F}\pm}({\bf r}) = \delta_{x,x_0} \frac{1}{\sqrt{L_y+1}} \left[
\begin{array}{c}
\sin (\pi y/2)\\
\mp \sin (\pi y/2)
\end{array}\right].
\end{eqnarray}
Note that the momentum-dependent phase shift $\theta_{\mysw{F}\pm}= -k/3$ cancels the relative geometric phase $k[\delta(y,B)-\delta(y,A)]$ leading to extremely simple local Wannier orbital at $x=x_0$. Repeatedly applying the lattice displacement operator $T_{x}$ on one Wannier orbital, all orbitals at different locations can be constructed. Since $[T_{x}, H_{t}] =0$, all the orbitals share the same energy and form a flat band. This is the one-dimensional analog of the Landau level degeneracy for two-dimensional electrons in magnetic field. The peculiar edge topology at nanoscale replaces the role of the magnetic field in 2D and quenches the kinetic energy of the carriers.

At first sight, it is rather counterintuitive that the local Wannier orbital cannot move around by quantum hopping. The static nature is due to perfect destructive quantum interferences which make hopping amplitudes from different sites cancel each other. Thus, the open boundaries of armchair shape play a crucial role for the birth of the Wannier orbitals. Furthermore, since the wave function is not zero only at odd $y$ coordinates (see Fig.~\ref{fig:Wannier}), it is clear that the adjacent orbitals have zero overlap. Thus, the Mielke-Tasaki mechanism does not work to couple neighboring orbitals magnetically.

Let us concentrate on the flat-band regimes $E= \pm t$ for the armchair GNR with odd $L_y$. Due to the particle-hole symmetry for the Hubbard model, the low-energy physics is dictated by the doping level disregarding whether it is electron or hole doped. Thus, it is convenient to introduce the positive-definite doping level $x_d \equiv |\langle n \rangle -1|$, where $\langle n \rangle$ is the average electron number per site. The lower and upper bounds of the doping rate $x_d$ for the flat-band regime are obtained from the Fermi momenta $k_m$ of the dispersive bands intersecting the flat band. For those dispersive bands, the Fermi momentum satisfies $\cos(k_m/2) + \cos(q_m) =0$, which leads to $k_{m} = 2\pi - 2q_m$, where $m=L_y, L_y-1,...,(L_y+3)/2$: there are $(L_y-1)/2$ pairs of Fermi points crossing the flat band. Thus, the system is in the flat-band regime for $x_{d,{\rm min}} < x_d < x_{d,{\rm max}}$, where 
\begin{eqnarray}
x_{d,{\rm min}}=\frac{1}{\pi L_y}\sum_m k_m=\frac14 - \frac{1}{4L_y},
\nonumber\\
x_{d,{\rm max}}=x_{d,{\rm min}} + \frac{1}{L_y} = \frac14+\frac{3}{4L_y}.
\end{eqnarray}
Therefore, the flat-band regime shrinks as the transverse width increases and eventually goes to zero in the two-dimansional limit. This trend highlights the importance of finite transverse width of the system and why the flat-band physics is no longer the dominant player in 2D. In the following, we will try to write down the effective field theory for both the flat and dispersive bands in weak coupling.

\section{Weak-Coupling Analysis}\label{sec:weakcoupling}

Even though we have derived the analytical form of wave functions in armchair GNR, it is still quite complicated to write down the effective field theory. In the flat-band regime, after integrating out gapped bands, there remains a flat band with intersecting dispersive bands. Note that the lattice fermion can be decomposed into eigenstates,
\begin{eqnarray}
c_\alpha(x,y,\Lambda) = \frac{1}{\sqrt{L_x}} \sum_{k} e^{ikx} \sum_{m,s} \phi_{ms}({\bf y}) c_{ms\alpha}(k),
\end{eqnarray}
where the extra geometric factor is included in the modified wave function $\phi_{ms}({\bf y}) = e^{ik\delta({\bf y})} \Phi_{ms}({\bf y})$. In the low-energy limit, one can approximate all intersecting bands with linear dispersions, the above expansion is then greatly simplified. 

Let us work out the chiral-field expansion in the flat-band regime at $E=t$ as an example. In that case, the effective low-energy theory is described by the flat band and the intersecting dispersive bands of the antibonding sector $s = 1$. Thus, the lattice fermion can be decomposed into the flat-band and pairs of chiral field operators,
\begin{eqnarray}
c_\alpha({\bf r}) \simeq \phi_{\mysw{F}}({\bf y}) \psi_{\mysw{F}\alpha}(x)
+ \sum_{m} \sum_{\mysw{P}} e^{i\mysw{P}k_{m}x} \phi_{\mysw{P}m}({\bf y}) \psi_{\mysw{P}m\alpha}(x),
\nonumber \\
\end{eqnarray}
where $P = R/L = +/-$ represents the chirality.
The modified (including the geometric phases) wave function for the flat-band orbital is
\begin{eqnarray}
\phi_{\mysw{F}}({\bf y}) = \frac{1}{\sqrt{L_y+1}} \left[
\begin{array}{c}
\sin (\pi y/2)\\
- \sin (\pi y/2)
\end{array}\right],
\label{eq:phiF}
\end{eqnarray}
and those for the right/left-moving plane waves at Fermi point $\pm k_m$ are 
\begin{eqnarray}
\phi_{\mysw{P}m}({\bf y}) = \frac{1}{\sqrt{L_y+1}}  \left[
\begin{array}{c}
 e^{i\mysw{P} k_{m}\delta_{\mysw{A}}} \sin (q_m y) \\
e^{i\mysw{P} k_m[2/3+\delta_{\mysw{B}}]} \sin (q_m y)
\end{array}
\right],
\label{eq:phiPm}
\end{eqnarray}
where the shorthand notation is introduced $\delta_{\mysw{A/B}} = \delta(y, A/B)$. Substituting the chiral-field expansion into the lattice Hamiltonian, one can easily derive the effective field theory in the flat-band regimes.

Since the density of states is divergent for the flat band, the dispersive bands can be dropped to the leading order approximation. The effective Hamiltonian, keeping the flat band only, is rather simple,
\begin{eqnarray}
H_{\mysw{F}} = U_{\mysw{F}} \sum_{x} n_{\mysw{F}\uparrow}(x) n_{\mysw{F}\downarrow}(x),
\end{eqnarray}
where $n_{\mysw{F}\alpha}(x)$ is the number density of each spin flavor and $U_{\mysw{F}}$ is the effective on-site interaction for the flat-band orbitals, which can be computed by projection onto the flat band,
\begin{eqnarray}
U_{\mysw{F}} = U \sum_{{\bf y}} |\phi_{\mysw{F}}({\bf y})|^2 |\phi_{\mysw{F}}({\bf y})|^2 = \frac{U}{L_y+1}.
\end{eqnarray}
Note that the kinetic energy is quenched in the flat band and thus the ground state contains no quantum fluctuations. To avoid the cost of $U_{\mysw{F}}$, the ground state consists of the maximum number of singly-occupied Wannier orbitals, which leads to local magnetic moments. Since there is no overlap between adjacent orbitals, these magnetic moments are free and give rise to a large ground-state degeneracy.

To lift the large degeneracy of the ground state, interaction with the itinerant carriers in the dispersive bands {\em must} be included. While the effective Hamiltonian for the interaction contains other terms, the terms to play a key role are the exchange interactions which couple the local moments and the itinerant carriers,
\begin{eqnarray}
H_{\mysw{J}} =  \sum_{x,m} -J_m\: {\bf S}_{\mysw{F}}(x) \cdot {\bf S}_{m}(x),
\label{ExchangeHamiltonian}
\end{eqnarray}
where ${\bf S}_{\mysw{F}}(x)$ is the spin density operator for the local moments and ${\bf S}_{m}(x) = {\bf S}_{\mysw{R}m}(x)+ {\bf S}_{\mysw{L}m}(x)$ is the spin density of itinerant carriers in the crossing band $m$. The exchange integral is given by,
\begin{eqnarray}
J_{\mysw{P}m} = 2 \sum_{{\bf y}, {\bf y'}} \phi^{*}_{\mysw{F}}({\bf y}) \phi^{}_{\mysw{F}}({\bf y'}) \phi_{\mysw{P}m}({\bf y}) \phi^{*}_{\mysw{P}m}({\bf y'}) V_{{\bf y},{\bf y'}}.
\end{eqnarray}
Using Eqs.\ (\ref{eq:phiF}) and (\ref{eq:phiPm}), we obtain the exchange integral due to the on-site interaction $V_{{\bf y},{\bf y'}}= \delta_{{\bf y}, {\bf y'}} U$; it has a rather simple form,
\begin{eqnarray}
J_{m} &=& 2 U \sum_{\bf y} |\phi_{F}({\bf y})|^2 |\phi_{\mysw{P}m}({\bf y})|^2 
\nonumber\\
&=& \frac{2U}{(L_y+1)^2}
\sum_{y=odd} 2 \sin^2 (q_m y)
= \frac{U}{L_y+1}.
\end{eqnarray}
The subscript $P=R/L$ is dropped because the coupling strength does not depend on the chirality. Thus, the on-site interaction induces a ferromagnetic coupling between the local moments in the flat band and the itinerant spins in the dispersive bands.

The exchange coupling $J_{m}$ tends to align the local moments from the flat band because it does not cause any extra kinetic energy. The ferromagnetically aligned moments act back on the itinerant carriers and induce finite polarization in the dispersive bands. The interacting Hamiltonian $H_{\mysw{F}}+ H_{\mysw{J}}$ therefore shows interesting two-step flat-band ferromagnetism -- electronic correlations in the flat band give rise to local moments without direct exchange coupling, while the presence of gapless itinerant carriers mediates the ferromagnetic order. The significant feature of the armchair GNR is that, even within the one-orbital Hubbard model without any magnetic impurity nor additional localized levels, the electronic correlations give rise to {\em both} local moments and itinerant carriers simultaneously due to the peculiar topology of the open edges.

It is also interesting to consider the finite-size effect arisen from the length $L_x$ of the armchair GNR. If one imposes the periodic boundary conditions along the $x$-direction, the system becomes a short segment of armchair nanotube. In this case, only when the quantized momenta $k_{x} = 2l\pi/L_{x}$ ($l=0,1,...,L_x-1$) coincide with the Fermi points $\pm k_{m}$, the gapless itinerant carriers are present and the ferromagnetic ground state is realized. For open boundary conditions with finite $L_x$, the situation is much more complicated; $k_{x}$ is no longer good quantum number and the band structure can be deformed by finite $L_x$. However, from numerical calculations for the tight-binding Hamiltonian $H_{t}$, we have found that the energy spectrum around the flat-band level $E = \pm t$ is not affected by finite $L_x$ and can be accurately approximated by the quantization rule $k_{x} = \tilde{l}\pi/(L_{x}+1)$ ($\tilde{l} = 1, 2, ..., L_x$). When the quantized momenta $k_{x}$ coincide with the Fermi points $k_{m}$, ferromagnetism sets in with the help of these gapless itinerant carriers. Therefore, depending on specific choice of $L_{x}$, the ground state of the armchair GNR in the flat-band regime can be ferromagnetic or Curie-like paramagnetic. 

\section{Non-Abelian Density Matrix Renormalization Group}\label{sec:DMRG}

\begin{figure}
\centering
\includegraphics[width=0.8\columnwidth]{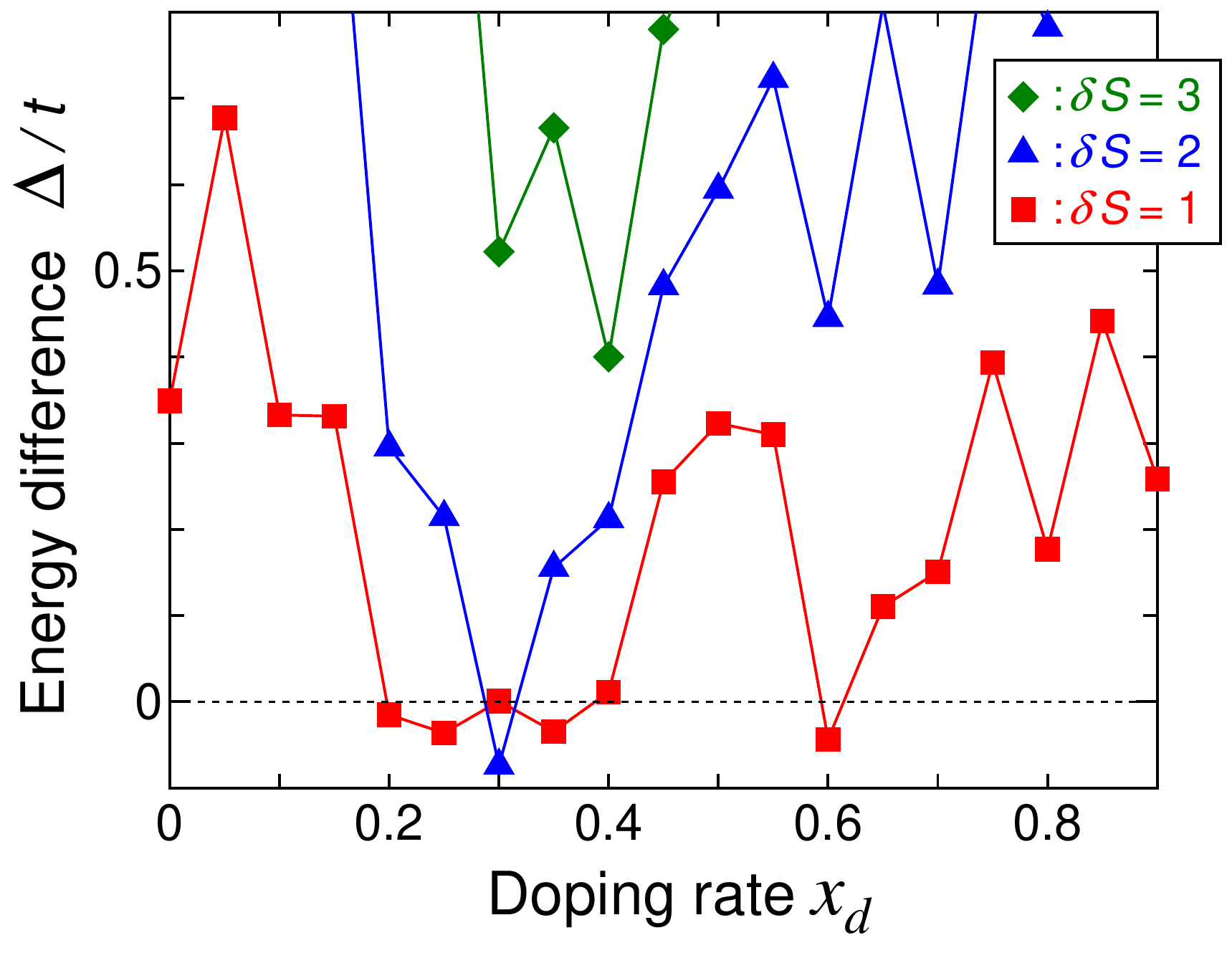}
\caption{\label{fig:GAP} (Color online) 
Doping dependence of the energy difference $\Delta(x_d, \delta S)$ between higher-spin and lowest-spin states in the armchair GNR with $L_x = 2$ and $L_y = 5$ for $U/t = 4$. The squares, triangles, and diamonds represent the data for $\delta S = S-S_0 = 1$, $2$, and $3$, respectively.}
\end{figure}

\begin{table}
\caption{
Energy difference per unit cell $\Delta(x_d)$ between the ferromagnetic ground state and the paramagnetic one in the armchair GNR with $L_x = 2$ and $L_y = 5$ for the flat-band regime. Here $S$ is the total spin of the ground state and the hopping amplitude is chosen $t = 3$ eV.
\label{tab:GAP}
}
\begin{ruledtabular}
\begin{tabular}{ccccc}
$x_d$ & $S$ & $U/t=2$   & $U/t=4$   & $U/t=8$   \\
\hline
  0.25   & 3/2 & -44 meV  & -54 meV  & -57 meV  \\
  0.30   & 2   & -115 meV & -111 meV & -87 meV \\
  0.35   & 3/2 & -48 meV  & -52 meV  & -34 meV  \\
\end{tabular}
\end{ruledtabular}
\end{table}

To check the validity of the weak-coupling scenario and see whether it survives for the realistic coupling regime, we choose the non-Abelian DMRG method.\cite{McCulloch02} It is important to emphasize that, to look for the higher-spin ground state, the non-Abelian approach is more powerful and convenient compared to the conventional DMRG\cite{White92,White93} because the former makes the full use of the spin SU(2) symmetry. Employing the non-Abelian DMRG method, we can compute the energy difference between the higher-spin (ferromagnetic) state and the lowest-spin (paramagnetic) state,
\begin{eqnarray}
\Delta (x_d, \delta S) \equiv E_0(x_d,S) - E_0(x_d,S_0),
\end{eqnarray}
where $E_0(x_d,S)$ is the lowest energy in the subspace with the doping rate $x_d$ and the total spin $S$. Furthermore, $\delta S \equiv S - S_0$, where $S_0=0,1/2$ denotes the lowest spin depending on whether the number of carriers is even or odd. The calculation is performed for the system with $L_y = 5$ and $L_x = 2$, for which the quantized momenta $k_x = \tilde{l} \pi/3$ coincide with the Fermi points of the dispersive bands and therefore the weak-coupling theory predicts the ferromagnetic ground state. The number of SU(2) multiplets kept is up to $450$, typically corresponding to $1000$-$3000$ U(1) states. The truncation error is of order $10^{-5}$ or less, and the results are extrapolated to the limit of zero truncation error. 

\begin{figure}
\centering
\includegraphics[width=\columnwidth]{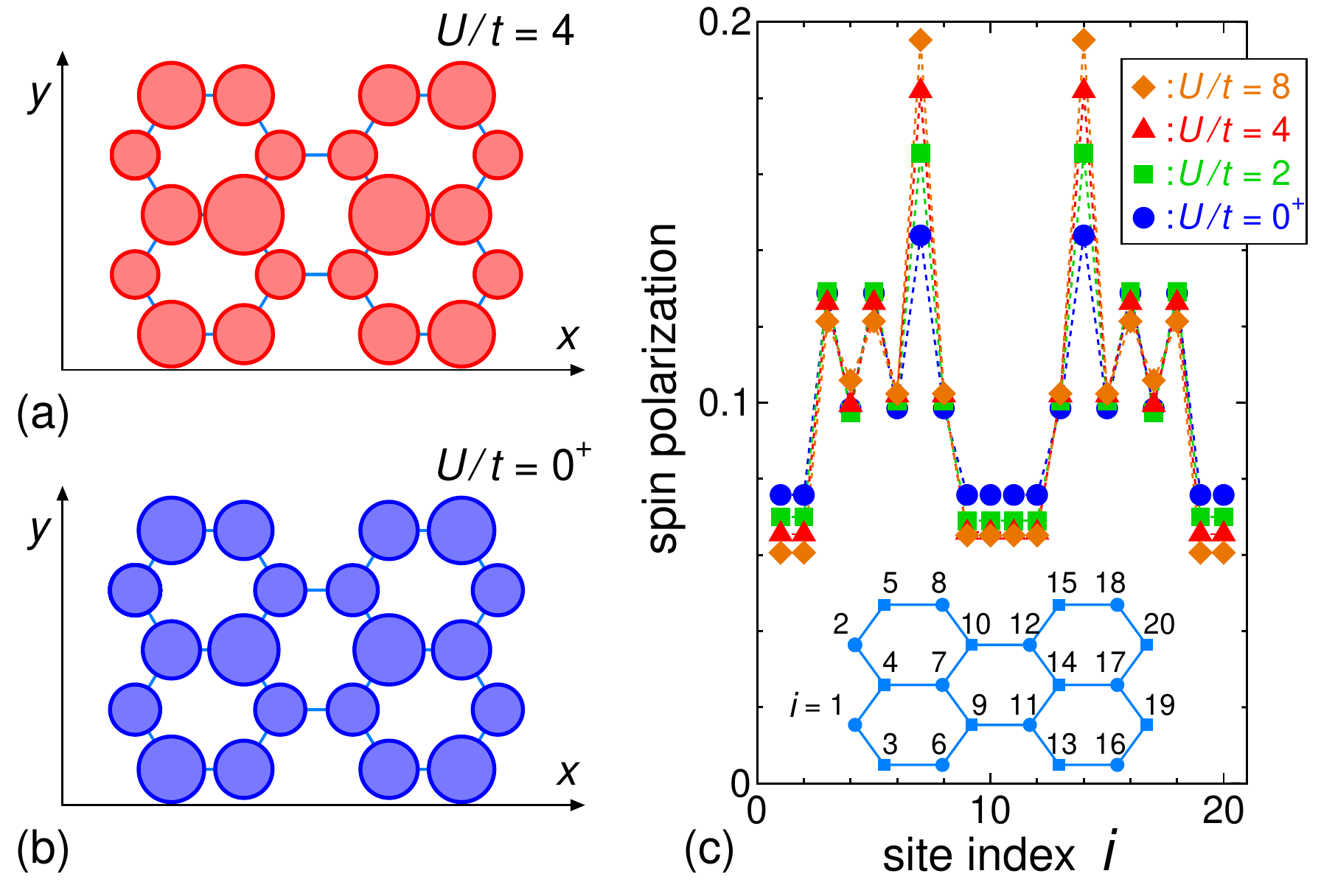}
\caption{\label{fig:WF2} (Color online) Spin polarization profile $\langle S^z({\bf r}) \rangle$ of the ferromagnetic ground state for the armchair GNR with $L_{x}=2$ and $L_{y}=5$. It is at the optimal doping $x_d^*=0.30$ with the interaction strength (a) $U/t = 4$. (b) $U/t = 0^+$. The values of $\langle S^z({\bf r}) \rangle$ are represented by the areas of the shaded circles. It is remarkable that the profiles with intermediate and weak interaction strength are almost identical. (c) The values of $\langle S^z({\bf r}) \rangle$ at each site for $U/t=0^+, 2, 4, 8$ represented by solid circles, squares, triangles and diamonds.}
\end{figure}

Figure~\ref{fig:GAP} shows the energy difference $\Delta (x_d, \delta S)$ as a function of the doping rate $x_d$ and the spin $\delta S$ for $U/t=4$. We do find numerically the ferromagnetic (higher-spin) ground state in the flat-band regime, $x_d=0.25, 0.30, 0.35$.\cite{FM-at-highdoping} The results with $U/t = 2, 8$ (not shown here) also show a similar doping-rate dependence, supporting the ferromagnetism for the flat-band regime.

Collecting all data for $\Delta (x_d, \delta S)$ together, we can determine the ground state at each doping level and its energy gain per unit cell, $\Delta (x_d)$, which is the energy difference between the higher-spin ferromagnetic ground state and the paramagnetic one with lowest spin,
\begin{eqnarray}
\Delta(x_d) = \frac{1}{L_x}[E_{0}(x_d) - E_{0}(x_d,S_{0})],
\end{eqnarray}
where $E_{0}(x_d)$ is the ground state energy at doping level $x_d$. The results for the flat-band regime ($0.2<x_d<0.4$ for $L_y=5$) and $U/t = 2, 4, 8$ are summarized in Table~\ref{tab:GAP}. The optimal doping occurs at $x_d^* = 0.3$ as predicted by the weak-coupling theory. Thus, the non-Abelian DMRG results support the carrier-mediated ferromagnetism predicted from the analytical approach in weak coupling even when the interaction strength is in the intermediate regime.

To further verify the role of the flat band, we also calculate the profile of spin density in the ground state.  Since the interaction strength is now larger than the hopping amplitude, one may guess any peculiar feature in the band structure should be suppressed. Figure~\ref{fig:WF2} shows the results at the optimal doping $x_d=0.3$ with total spin $S = 2$. Remarkably, the spin polarization for finite $U/t=4$ has a similar profile to that in the weak-coupling limit $U \to 0^+$ obtained from the eigen-wavefunctions of the tight-binding model $H_{t}$. The result clearly indicates that the flat-band orbitals still play a significant role in the ferromagnetic ground state for finite $U/t$.
The physical picture developed in weak coupling thus applies rather well and extends smoothly to the intermediate- and strong-coupling regime.

As the momentum $k_x$ is discretized in the system with finite $L_x$, 
it is important to see how the properties 
of the system change depending on $L_x$. 
For armchair GNR, the large unit cell and peculiar nature of 
the flat-band states lead to slow convergence of the DMRG calculation 
and make it difficult to treat the system with larger $L_x$, 
unfortunately.\cite{slow-convergence} 
Nevertheless, we have performed numerical calculation for other 
one-dimensional-lattice models which have essentially 
the same band structure as that of armchair GNR. We have then found that the models with larger number of unit cells 
indeed exhibits the itinerant ferromagnetism with the properties 
expected from the weak-coupling theory in Sec.\ \ref{sec:weakcoupling}, 
even including the peculiar finite-size effect.
The results will be reported elsewhere.\cite{HikiharaLM} 
Furthermore, we also emphasize that, 
as we will see in Sec.\ \ref{sec:LSDA}, the first-principles 
calculation for infinite-length armchair GNR also shows the itinerant 
ferromagnetism for the flat-band regime, in accordance with the 
weak-coupling prediction. 
These observations support that the carrier-mediated ferromagnetism 
found in this section 
would survive for larger $L_x$ and connect to the thermodynamic limit $L_x \to \infty$.

\section{Local Spin Density Approximation}\label{sec:LSDA}

The complimentary approaches of the weak-coupling analysis and the non-Abelian DMRG method establish the ferromagnetic ground state in armchair GNR within the Hubbard-type model. To treat GNRs, however, one must take account of the realistic band structure beyond the tight-binding approximation as well as the effect of unscreened Coulomb interactions. For the purpose, we have carried out first-principles calculation of the local spin-density approximation within density functional theory. The self-consistent band structure calculations under lattice optimization were performed using the full-potential projected augmented wave method \cite{Blochl94,Kresse99} as implemented in the VASP package.\cite{Kresse93,Kresse96} 

\begin{figure}
\centering
\includegraphics[width=\columnwidth]{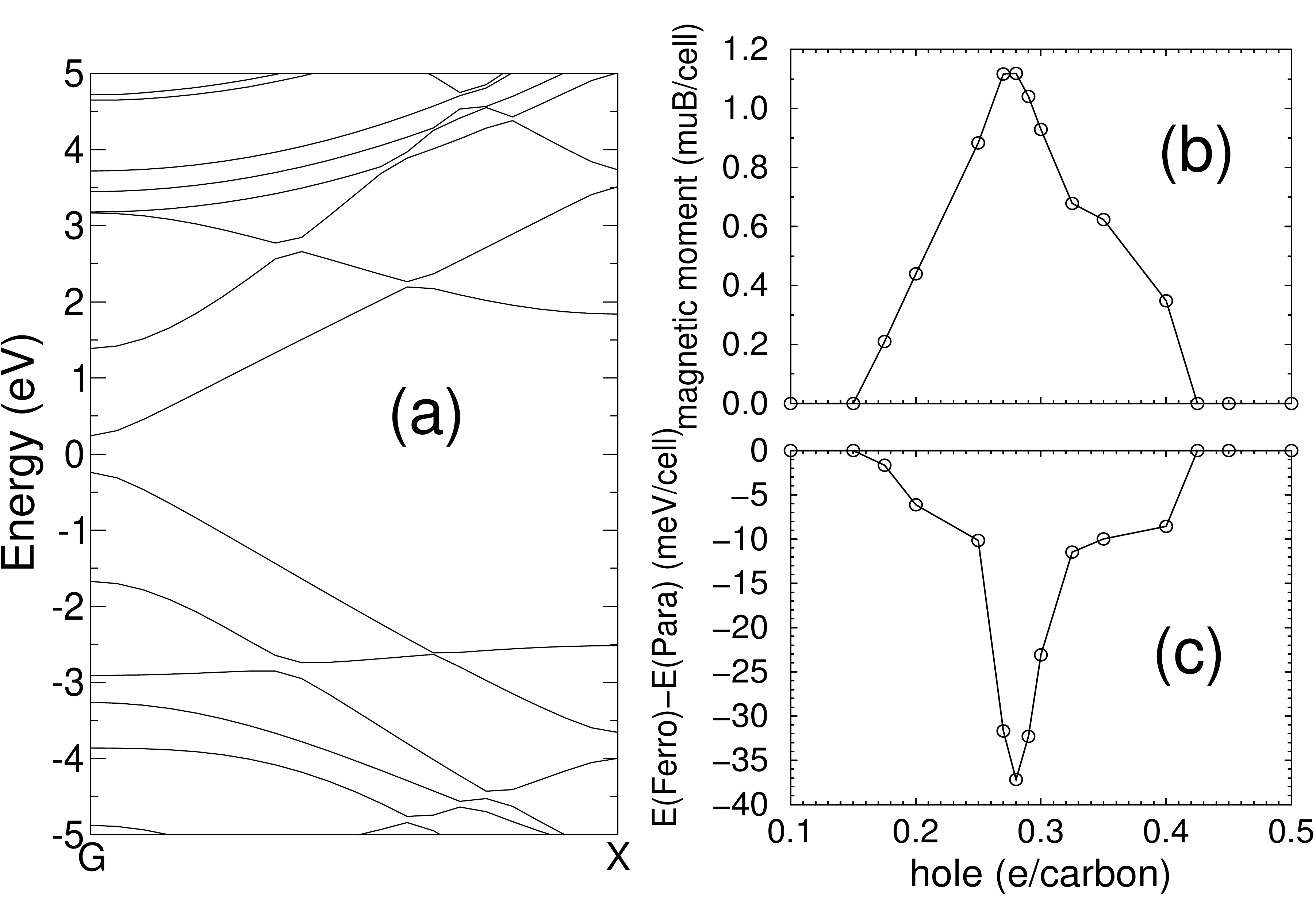}
\caption{\label{bsnm} (a) Band structure of the undoped, infinitely long armchair GNR with $L_{y}=5$. (b) Doping dependence of the spin polarization per unit cell. (c) Energy difference per unit cell $\Delta (x_h)$ between ferromagnetic and paramagnetic states as a function of the hole doping.}
\end{figure}

Figure \ref{bsnm}(a) shows our result of the band structure of the armchair GNR with $L_{y}=5$ at half filling. The numerics show that the band structure of the undoped armchair GNR is more or less similar to the tight-binding results. There exists a narrow band with bandwidth $\sim 0.4$ eV located at $2.5 \sim 2.9$ eV below the Fermi level crossed by two itinerant bands. Furthermore, one can find similar structure in the unoccupied counterpart though the particle-hole symmetry does not hold for this case.

\begin{figure}
\centering
\includegraphics[width=\columnwidth]{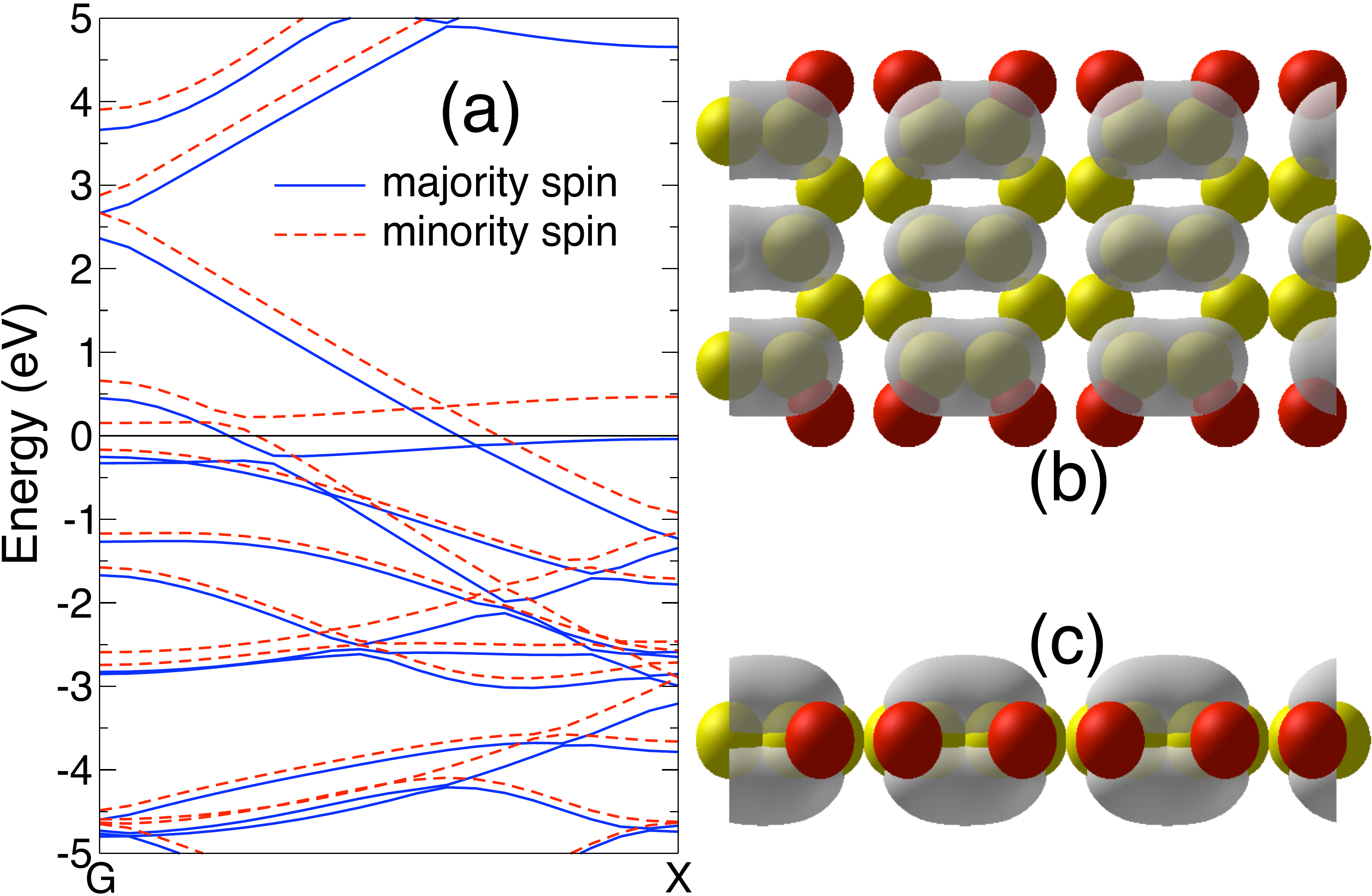}
\caption{\label{bsfm} (Color online) (a) Spin decomposed band structure of the infinitely long GNR with $L_y=5$ at the optimal hole doping $x_h^* \simeq 0.28$. (b) Top and (c) side views of the spin density distribution (shown in grey color). The yellow (light) and red (dark) spheres denote C and H atoms respectively.}
\end{figure}

Although the lower ``flat band" has a finite bandwidth ($\sim 0.4$ eV) and thus is not flat anymore, it is still massive enough compared with the other two dispersive bands. As a result, the two-step carrier-mediated ferromagnetism still works as will be described below. Figure \ref{bsnm}(b) presents the magnetic moment per unit cell at different hole doping levels $x_h = 1-\langle n \rangle$. Upon hole doping, the magnetic moment goes up and reaches its maximum, $\sim 1.1 \mu_B$ per unit cell, at the optimal hole doping $x_h^* \sim 0.28$. In the mean time, as shown in Fig.\ \ref{bsnm}(c), the energy gain of the ferromagnetic state per unit cell raises noticeably, especially near the optimal doping, up to $\sim $ 37 meV/cell. These results clearly support the emergence of the ferromagnetism.

The obtained value of the magnetic moment $\sim 1.1 \mu_B$ per unit cell indicates that the nearly-flat band is almost fully-polarized at the optimal doping. This feature is also shown by the spin decomposed band structure in Fig. \ref{bsfm}(a). The bandwidth of the nearly-flat band at the optimal doping is slightly suppressed to $\sim 0.3$ eV because of the reduced density. The exchange splitting between opposite spins is about 0.5 eV. Figures~\ref{bsfm}(b) and (c) demonstrate respectively the top and side views of the spin density distribution at the optimal doping. It is truly remarkable that the profile of the spin density is almost identical to the Wannier orbital in the weak-coupling limit (see Fig.~\ref{fig:Wannier}). The nodes caused by destructive quantum interferences can be seen clearly in the numerics. We emphasize that not only the dumb-bell shape of the spin density emerges as predicted, the optimal doping concentration and the size of the spin polarization realized also agree with the prediction of the weak-coupling theory. 

As the hole doping increases further, the magnetic moment as well as the energy gain of the ferromagnetic state are suppressed. The ground state eventually turns nonmagnetic at larger hole doping.
It is interesting to notice that the ferromagnetic ground state, as shown in Fig.~\ref{bsnm}, exists for $0.15 \leq x_h \leq 0.42$, which is wider than the flat-band regime between $x_{\rm min}=0.2$ and $x_{\rm max}=0.4$ predicted by weak-coupling theory for $L_y=5$ armchair GNR. The ferromagnetic phase obtained by the local spin-density approximation then seems to persist even slightly outside the flat-band regime. This shall not be surprising since the interaction strength is no longer weak in comparison with the kinetic energy. To address this issue, one can calculate the variational energy of the Hubbard model in the armchair GNR\cite{Lee08} to pin down the ferromagnetic regime at finite interaction strength $U$. Considering the GNR slightly outside the flat-band regime, although costing higher kinetic energy, it is still preferential to fill in particles/holes in the flat band since the exchange energy is lowered. Thus, we expect that the ferromagnetic ground state can exist even outside the flat-band regime. The variational calculations indeed show that the above understanding is correct and the ferromagnetic phase exists in a wider range than the flat-band regime. Therefore, it is expected that the inclusion of the realistic Coulomb interaction will give rise to similar enhancement of the flat-band ferromagnetism as demonstrated in Fig.~\ref{bsnm} here.

\section{Discussions and Conclusions}\label{sec:conclusion}

Here we would like to elaborate on several subtle points which were not discussed in previous sections. First, we note that the weak-coupling theory in Sec.\ \ref{sec:weakcoupling} developed for the Hubbard model in the armchair GNR is pretty robust against lattice distortions. In practical graphite network, the hopping along horizontal and tilted vertical bonds is expected to be slightly different. Though the deviation shifts the flat band from $E=\pm t$, the bands remain flat and the same mechanism of the ferromagnetism applies.

Furthermore, the profile of the short-range interaction does not seem to do much harm either. Following steps similar to those in Sec.\ \ref{sec:weakcoupling}, we can compute the exchange coupling due to nearest-neighbor interaction $V_{\perp}$ (tilted vertical bonds) and $V_{\parallel}$ (horizontal bonds). Since the product of the flat-band wave function $\phi_{\mysw{F}}({\bf y})\phi_{\mysw{F}}({\bf y'})=0$ if the coordinates are connected by a tilted vertical bond, $V_{\perp}$ does not give rise to any exchange coupling. On the other hand, the nearest-neighbor interaction along the horizontal bonds $V_{\parallel}$ gives rise to non-vanishing exchange coupling,
\begin{eqnarray}
J^{\mysw{V}_{\parallel}}_{m} &=& 2 V_{\parallel}  \sum_{\bf y} \phi^*_{\mysw{F}}({\bf y}) \phi_{\mysw{F}}(\overline{\bf y})
\phi^*_{\mysw{P}m}(\overline{\bf y}) \phi_{\mysw{P}m}({\bf y})
\nonumber\\
&=& \frac{2V_{\parallel}}{(L_y+1)^2} (-2\cos k_m) \sum_{y=odd} \sin^2 (q_m y)
\nonumber\\
&=& - \frac{V_{\parallel}}{L_y+1} \cos k_m.
\end{eqnarray}
Here we have used the expression for the interaction $V_{{\bf y},{\bf y'}} = \delta_{\overline{\bf y}, {\bf y'}} V_{\parallel}$ with the notation $\overline{\bf y} = (y, B/A)$ for ${\bf y} = (y, A/B)$. As expected, the exchange couplings arisen from right-/left-moving fields in the same dispersive band are the same and thus the subscript $P=R/L$ is dropped. We thereby find that, in the presence of the nearest-neighbor interaction $V_{\perp}$ and $V_{\parallel}$, the exchange coupling in Eq.~(\ref{ExchangeHamiltonian}) becomes
\begin{eqnarray}
J_{m} = \frac{1}{L_y+1} (U-V_{\parallel} \cos k_m).
\end{eqnarray}
Since $V_{\perp}, V_{\parallel} <U$ is often expected, the exchange coupling is still ferromagnetic and the picture does not change.

The above calculations can be generalized to the screened short-ranged interaction. Suppose the spatial profile of the screened interaction goes as $\exp(-x/\xi)/\sqrt{x^2+l_0^2} $, where $l_0$ is a short-range cutoff (comparable to the lattice constant) and $\xi$ is the length scale of the short-ranged interaction. Ignoring the complicated form factor due to detail orbital overlapping, the exchange coupling takes the general form $J_m(x) \sim \exp(-x/\xi) \cos (k_m x) /x$. As expected, the exchange coupling decreases as the distance is far apart. Furthermore, the oscillatory factor $\cos (k_m x)$ makes the couplings to the dispersive bands with different signs and tends to cancel each other. It will further suppress the effects of the exchange coupling beyond the nearest neighbors. This trend is in agreement with our first-principles calculations where the true long-ranged Coulomb interaction is included.

To realize the flat-band ferromagnetism in armchair GNR, the crucial challenge lies in how to achieve the appropriate doping level. One of graphene's superior properties is its pronounced ambipolar electric field effect.\cite{Novoselov04,Novoselov05,Zhang05} By applying gate voltages, the charge carriers can be tuned between electrons and holes with concentration up to $10^{13}$ cm$^{-2}$. Even so, it is unlikely that the external gate voltage alone can pour enough electrons/holes into the system to reach the flat-band regime. Another route to dope GNR is via chemical doping. It was demonstrated\cite{Ohta06} that the chemical dopants in the substrate can markably change the carrier density. Perhaps the combination of both methods can be even more efficient.

In conclusion, by combining the weak-coupling analysis, the non-Abelian DMRG technique, and the first-principles calculations, we show how ferromagnetism occurs in armchair GNR -- electronic correlations give rise to magnetic moments in the flat band and the itinerant carriers in the dispersive bands mediate ferromagnetic coupling between these uncoupled moments. Recently, there are proposals\cite{Rycerz07,Trauzettel07} to use GNR to build transistors and spin qubits. While these proposals take care of many realistic issues, the electronic correlations are ignored. Our study here show that electronic correlations in GNR can bring up surprises such as the carrier-mediated flat-band ferromagnetism. Therefore, it is crucially important to include the correlation effects when we try to realize these proposals into devices.

We acknowledge Leon Balents, Greg Fiete, Yukitoshi Motome and Tsutomu Momoi for valuable discussions and comments. HHL, HTJ, BLH and CYM appreciates financial supports from National Science Council in Taiwan. The hospitality of KITP in Santa Barbara is also greatly appreciated.


\begin{thebibliography}{10}
\bibitem{FM-at-highdoping}
The high-spin ground state with $\delta S = 1$ for $x_d = 0.6$ 
seen in Fig.\ \ref{fig:GAP} is due to an intersection of 
two dispersive bands at a high-doping rate [see Fig.\ \ref{fig:Wannier} (b)] 
and does not lead to a bulk ferromagnetism in the thermodynamic limit.
\bibitem{slow-convergence}
In the lattice geometry of armchair GNR, rather large unit cells consisting 
of 2$L_y$ sites are connected via $(L_y-1)/2$ bonds, resulting in 
a rapid increase of the number of sites with $L_x$ and a loose connection 
between unit cells. 
Furthermore, for the flat-band regime, there are many low-lying states 
with extremely small energy scale, especially for the cases of $L_x$ 
where the itinerant electron at the flat-band level is absent.
These facts lead to a very slow convergence of the DMRG calculation and 
require a large number of kept DMRG states for achieving accurate results.
\end{thebibliography}
\end{document}